\documentstyle[preprint,12pt,aps]{revtex}

\tightenlines
\begin{document}
\title{Creation of scalar particles in the presence of a constant electric field in
an anisotropic cosmological universe}
\author{V\'{\i}ctor M. Villalba \thanks{%
e-mail:villalba@ivic.ivic.ve}}
\address{Centro de F\'{\i}sica\\
Instituto Venezolano de Investigaciones Cient\'{\i}ficas, IVIC\\
Apdo 21827, Caracas 1020-A, Venezuela}
\maketitle

\begin{abstract}
In the present article we analyze the phenomenon of particle creation in a
cosmological anisotropic universe when a constant electric field is present.
We compute, via the Bogoliubov transformations, the density number of scalar
particles created.
\end{abstract}

\pacs{11.10, 03.65}

Quantum field theory in curved space-time is perhaps one of the most
interesting and puzzling problems in contemporary theoretical physics. After
the publication of the pioneer article by Hawking\cite{Hawking} about pair
production in the vicinity of a Schwarzschild black hole, a great body of
papers have been published, mainly trying to understand the mechanism that
gives origin to the thermal particle distribution and its relation to
thermodynamics. It is noteworthy that Hawking's result was preceded by a
series of articles where the question was to discuss particle production in
cosmological universes \cite{Parker,Zeldovich}. Almost all of the articles
published in this area deal with isotropic and homogeneous gravitational
backgrounds, mainly in deSitter and Robertson Walker models, and only a few
try to discuss quantum processes in anisotropic Universes.

The study of quantum effects in gravitational backgrounds with initial
singularities presents an additional difficulty. The techniques commonly
applied in order to define particle states are based on the existence of a
timelike Killing vector or an asymptotically static metric \cite{Birrel}. A
different approach is needed to circumvent the problem related to the
initial singularity. In this direction, the Feynman path-integral method has
been applied to the quantization of a scalar field moving in the The
Chitre-Hartle Universe \cite{Chitre,Fischetti}. This model has a curvature
singularity at $t=0$, and it is perhaps the best known example where a time
singularity appears and consequently any adiabatic prescription in order to
define particle states fails. A spin $1/2$ extension has been considered by
Sahni.\cite{Sahni}

A different approach to the problem of classifying single particle states on
curved spaces, is based on the idea of diagonalizing the Hamiltonian. This
technique permits one to compute the mean number of particles produced by a
singular cosmological model, and in particular by the Chitre-Hartle Universe 
\cite{Chitre}.

An interesting scenario for discussing particle creation processes is the
anisotropic universe associated with the metric

\begin{equation}
ds^{2}=-dt^{2}+t^{2}(dx^{2}+dy^{2})+dz^{2}.  \label{a1}
\end{equation}
The line element (\ref{a1}) presents a space-like singularity at $t=0.$ The
scalar curvature is $R=2/t^{2}$, and consequently, the adiabatic approach 
\cite{Birrel}cannot be applied in order to define particle states. With the
help of the Hamiltonian diagonalization method \cite{Grib,bukh2,bukh3},
Bukhbinder \cite{Bukhbinder} has been able to compute the rate of scalar
particles produced in the space with the metric (\ref{a1}), obtaining as
result a Bose-Einstein distribution. More recently \cite{villalba}, a
quasiclassical approach has been applied to compute the rate of scalar as
well as Dirac particles in the metric (\ref{a1}).

The introduction of an external electric field permits one to consider an
additional source of quantum processes. The density of particles created by
an intense electric field was first calculated by Schwinger\cite{Schwinger},
different authors\cite{Grib,Nikishov} have discussed this problem. Pair
creation of scalar particles by a constant electric field in a 2+1 de Sitter
cosmological universe has been analyzed by Garriga\cite{Garriga}. Quantum
effects associated with scalar and spinor particles in a quasi-Euclidean
cosmological model with a constant electric field are discussed by
Bukhbinder and Odintsov\cite{Odintsov}. It is the purpose of the present
article to compute, via the quasiclassical approach \cite
{Costa,villalba1,villalba2}, the density of scalar particles created in the
background field (\ref{a1}) when a constant electric field is present. The
idea behind the method is the following: First, we solve the relativistic
Hamilton-Jacobi equation and, looking at its solutions, we identify positive
and negative frequency modes. Second, we solve the Klein Gordon-equation
and, after comparing with the results obtained for the quasiclassical limit,
we identify the positive and negative frequency states. This technique has
already been successfully applied in different scenarios\cite
{Costa,villalba1,villalba2}.

The relativistic Hamilton-Jacobi equation can be written as 
\begin{equation}
g^{\alpha \beta }(\frac{\partial S}{\partial x^{\alpha }}-eA_{\alpha })(%
\frac{\partial S}{\partial x^{\beta }}-eA_{\beta })+m^{2}=0,  \label{2}
\end{equation}
where here and elsewhere we adopt the convention $c=1$ and $\hbar =1.$

The vector potential $A_{\alpha }$ 
\begin{equation}
A_{\alpha }=(0,0,0-Et),
\end{equation}
corresponds to a constant electric field $E\hat{k}$. The corresponding
invariants $F^{\mu \upsilon }F_{\mu \nu }=-2E^{2}$ and $F^{\mu \upsilon \ast
}F_{\mu \nu }=0$ indicate that there is no magnetic field. Since the metric $%
g_{\alpha \beta }$ associated with the line element (\ref{a1}) only depends
on $t,$ the function $S$ can be separated as 
\begin{equation}
S=F(t)+k_{x}x+k_{y}y+k_{z}z.  \label{3}
\end{equation}
Substituting (\ref{3}) into (\ref{2}) we obtain 
\begin{equation}
\dot{F}^{2}=\frac{k_{x}^{2}+k_{y}^{2}}{t^{2}}+(k_{z}+eEt)^{2}+m^{2}
\label{4}
\end{equation}
the solution of Eq. (\ref{4}) presents the following asymptotic behavior: 
\begin{equation}
\lim_{t\rightarrow \infty }F=\pm \frac{1}{2}t\sqrt{e^{2}E^{2}t^{2}-m^{2}}\mp 
\frac{m^{2}}{2eE}\log (eEt+\sqrt{e^{2}E^{2}t^{2}-m^{2}}),  \label{5}
\end{equation}
\begin{equation}
\Phi =e^{iS}\rightarrow Ce^{\pm \frac{i}{2}eEt^{2}}(eEt)^{\mp \frac{im^{2}}{%
2eE}}  \label{fi}
\end{equation}
as $t\rightarrow \infty ,$ and 
\begin{equation}
\lim_{t\rightarrow 0}F=\pm \sqrt{(k_{x}^{2}+k_{y}^{2})\log t,}\ \Phi
=e^{iS}\rightarrow Ct^{\pm i\sqrt{k_{x}^{2}+k_{y}^{2}}},\   \label{6}
\end{equation}
as $t\rightarrow 0,$ that is, in the initial singularity. Notice that the
time dependence of the relativistic wave function is obtained via the
exponential operation $\Phi \rightarrow \exp (iS).$ Here it is worth
mentioning that the behavior of positive and negative frequency states is
selected depending on the sign of the operator $i\partial _{t}.$ Positive
frequency modes will have positive eigenvalues and for negative frequency
states we will have the opposite. Then in Eq. (\ref{5}) and (\ref{6},\ref{fi}%
) upper signs are associated with negative frequency values and the lower
signs correspond to positive frequency states. After making this
identification we can analyze the solutions of the Klein-Gordon equation in
the background field (\ref{a1}).

The covariant generalization of the Klein-Gordon equation takes the form 
\begin{equation}
g^{\alpha \beta }(\nabla _{\alpha }-ieA_{\alpha })(\nabla _{\beta
}-ieA_{\beta })\Phi -(m^{2}+\xi R)\Phi =0,  \label{7}
\end{equation}
where $\nabla _{\alpha }$ is the covariant derivative, $R$ is the scalar
curvature, and $\xi $ is a dimensionless coupling constant which takes the
value $\xi =1/6$ in the conformal case, and $\xi =0$ when a minimal coupling
is considered. Substituting (\ref{a1}) into (\ref{7}) we obtain 
\begin{equation}
\frac{\partial ^{2}\Phi }{\partial t^{2}}-\frac{\partial ^{2}\Phi }{\partial
z^{2}}-2eEt\frac{\partial \Phi }{\partial t}-e^{2}E^{2}t^{2}\Phi -\frac{1}{%
t^{2}}\left( \frac{\partial ^{2}\Phi }{\partial x^{2}}+\frac{\partial
^{2}\Phi }{\partial y^{2}}\right) +(m^{2}+\frac{2\xi }{t^{2}})\Phi =0.
\label{8}
\end{equation}
Since eq. (\ref{8}) commutes with the linear momentum $\vec{p}=(-i\partial
_{x},-i\partial _{y}-i\partial _{z}),$ we have that the substitution 
\begin{equation}
\Phi =t^{-1}\Delta (t)e^{i(k_{x}x+k_{y}y+k_{z}z)},  \label{9}
\end{equation}
reduces eq (\ref{8}) to the ordinary second order differential equation. 
\begin{equation}
\frac{d^{2}\Delta }{dt^{2}}+\left( \frac{1}{t^{2}}\left(
k_{x}^{2}+k_{y}^{2}+2\xi \right)
+t^{2}e^{2}E^{2}+2tk_{z}E+k_{z}^{2}+m^{2}\right) \Delta =0  \label{10}
\end{equation}
whose solution, for $k_{z}=0$, can be expressed in terms of Whittaker
functions $M_{k,\mu }(z)$ and $W_{k,\mu }(z)$ \cite{Lebedev,Abramowitz} 
\begin{equation}
\Delta =z^{-1/4}(C_{1}M_{k,\mu }(z)+C_{2}W_{k,\mu }(z)),  \label{Delta}
\end{equation}
where $k,\mu $ and $z$ are given by the expressions 
\begin{equation}
z=ieEt^{2},\quad k=-i\frac{m^{2}}{4eE},\quad \mu =\frac{i}{4}\sqrt{4(2\xi
+k_{y}^{2}+k_{x}^{2})-1}.
\end{equation}
Looking at the asymptotic behavior of $M_{k,\mu }(z)$ and $W_{k,\mu }(z)$ as 
$\left| z\right| \rightarrow \infty $%
\begin{equation}
W_{k,\mu }(z)\sim e^{-z/2}z^{k},  \label{Whi}
\end{equation}
and as $z\rightarrow 0$%
\begin{equation}
M_{k,\mu }(z)\sim e^{-z/2}z^{1/2+\mu },  \label{Mi}
\end{equation}
we obtain that the solution (\ref{Delta}) having the asymptotic behavior
given by (\ref{5}) and (\ref{fi}) is 
\begin{equation}
\Delta _{\infty }^{+}=C_{\infty }^{+}z^{-1/4}W_{k,\mu }(z),\quad \Delta
_{\infty }^{-}=C_{\infty }^{-}z^{-1/4}W_{-k,\mu }(-z),  \label{delta1}
\end{equation}
where $C_{\infty }^{+}$ and $C_{\infty }^{-}$ are normalization constants

Analogously, we have that in the vicinity of the singularity, looking at the
quasiclassical solutions at $t=0$ (\ref{6}) the corresponding negative ``$-$%
'' and positive ``$+$'' frequency solutions take the form 
\begin{equation}
\Delta _{0}^{-}=C_{0}^{-}z^{-1/4}M_{k,\mu }(z),\quad \Delta
_{0}^{+}=C_{0}^{+}z^{-1/4}M_{k,-\mu }(z),  \label{delta2}
\end{equation}
where $C_{0}^{-}$ and $C_{0}^{+}$ are normalization constants, and the
function $M_{k,\mu }(z)$ can be expressed in terms of the Kummer
hypergeometric function $M(a,b,z)$ as follows: 
\begin{equation}
M_{k,\mu }(z)=e^{-z/2}z^{1/2+\mu }M(\frac{1}{2}+\mu -k,1+2\mu ,z).
\end{equation}
The Whittaker function $M_{k,\mu }(z)$ can be expressed in terms of $%
W_{k,\mu }(z)$ as\cite{Gradshteyn} 
\begin{equation}
M_{k,\mu }(z)=\frac{\Gamma (2\mu +1)}{\Gamma (\mu -k+\frac{1}{2})}e^{-i\pi
k}W_{-k,\mu }(-z)+\frac{\Gamma (2\mu +1)}{\Gamma (\mu +k+\frac{1}{2})}%
e^{-i\pi (k-\mu -\frac{1}{2})}W_{k,\mu }(z).  \label{recu}
\end{equation}
Using the above expression (\ref{recu}) we have that the negative frequency
solution $\Delta _{0}^{-}$ can be written in terms of $\Delta _{\infty }^{+}$
and $(\Delta _{\infty }^{-})$ as follows 
\begin{equation}
\Delta _{0}^{-}=\frac{\Gamma (2\mu +1)}{\Gamma (\mu -k+\frac{1}{2})}e^{-i\pi
k}\Delta _{\infty }^{-}+\frac{\Gamma (2\mu +1)}{\Gamma (\frac{1}{2}+\mu +k)}%
(-1)^{-1/4}e^{-i\pi \mu (k-\mu -\frac{1}{2})}(\Delta _{\infty }^{-})^{\ast }
\label{relation}
\end{equation}
where we have made use of the property $W_{-k,\mu }(-z)=(W_{k,\mu
}(z))^{\ast }$

\bigskip Since we have been able to obtain single particle states for in the
vicinity of $t=0$ as well as in the asymptote $t\rightarrow \infty $, we can
compute the density of particles created by the gravitational field. With
the help of the Bogoliubov coefficients \cite{Birrel,Grib}. From (\ref
{relation}) and the fact that $\Delta _{0}^{-}=\alpha \Delta _{\infty
}^{-}+\beta (\Delta _{\infty }^{-})^{\ast }$ we obtain 
\begin{equation}
\frac{\left| \beta \right| ^{2}}{\left| \alpha \right| ^{2}}=e^{2i\pi \mu }%
\frac{\left| \Gamma (\frac{1}{2}+\mu -k)\right| ^{2}}{\left| \Gamma (\frac{1%
}{2}+\mu +k)\right| ^{2}}.  \label{beta}
\end{equation}
Substituting into (\ref{beta}) the values for $\mu $ and $k$ we obtain 
\begin{equation}
\frac{\left| \beta \right| ^{2}}{\left| \alpha \right| ^{2}}=\frac{\cosh (%
\frac{\pi }{4}\sqrt{4(2\xi +k_{y}^{2}+k_{x}^{2})-1}-\frac{\pi m^{2}}{4eE})}{%
\cosh (\frac{\pi }{4}\sqrt{4(2\xi +k_{y}^{2}+k_{x}^{2})-1}+\frac{\pi m^{2}}{%
4eE})}e^{-\frac{\pi }{2}\sqrt{4(2\xi +k_{y}^{2}+k_{x}^{2})-1}}  \label{beta2}
\end{equation}
where we have used the relation\cite{Lebedev} 
\begin{equation}
\left| \Gamma (\frac{1}{2}+iy)\right| ^{2}=\frac{\pi }{\cosh \pi y}.
\end{equation}
The computation of the density of particles created is straightforward from (%
\ref{beta2}) and the normalization condition \cite{Mishima} of the wave
function 
\begin{equation}
\left| \alpha \right| ^{2}-\left| \beta \right| ^{2}=1,  \label{norma}
\end{equation}
then 
\[
n=\left| \beta \right| ^{2}=\left[ \left( \frac{\left| \beta \right| ^{2}}{%
\left| \alpha \right| ^{2}}\right) ^{-1}-1\right] ^{-1}.
\]
It should be noticed that, thanks to the normalization condition, we did not
have to compute the normalization constants $C$ appearing in the definition
of the single mode solutions (\ref{delta1},\ref{delta2}). Let us analyze the
asymptotic behavior of \ (\ref{beta2}) when the electric field vanishes.
Taking into account that $cosh(z)\sim e^{\left| z\right| }/2$ as $%
z\rightarrow \infty $, we readily obtain 
\begin{equation}
n\sim \frac{\left| \beta \right| ^{2}}{\left| \alpha \right| ^{2}}=\exp
(-\pi \sqrt{4(2\xi +k_{y}^{2}+k_{x}^{2})-1}),  \label{espectro}
\end{equation}
which is the result obtained in \cite{villalba}. \ Expression (\ref{espectro}%
) \ corresponds to a two dimensional Bose-Einstein thermal distribution with
an effective mass which differs from the value of $m$ appearing in Eq. (\ref
{7}). In the case of strong electric fields the density number of scalar
particles created takes the form 
\begin{equation}
n\approx \frac{\left| \beta \right| ^{2}}{\left| \alpha \right| ^{2}}=\exp (-%
\frac{\pi }{2}\tanh (\frac{\pi }{4}\sqrt{4(2\xi +k_{y}^{2}+k_{x}^{2})-1})%
\frac{m^{2}}{eE}-\frac{\pi }{2}\sqrt{4(2\xi +k_{y}^{2}+k_{x}^{2})-1}),
\label{electrico}
\end{equation}
showing \ that the density of particles created by the cosmological
background and the electric field \ (\ref{electrico}) is\ a Bose-Einsten
distribution with \ a chemical potential proportional to $\frac{m^{2}}{eE}.$
Integrating the particle density $n$ \ (\ref{electrico}) on momentum we
obtain the total number of particles created per unit volume. 
\begin{equation}
N=\frac{1}{V}\int ndk_{x}dk_{y}dk_{z}=\frac{1}{t^{2}(2\pi )^{2}}\int
nk_{\perp }dk_{\perp }dk_{z}.  \label{ene}
\end{equation}
In order to carry out the integration we have to notice that $n$ does not
depend on $k_{z}$ and consequently integration on $k_{z}$ is equivalent to
the substitution\cite{Grib,Nikishov} \ $\int dk_{z}\rightarrow eET$ where $T$
is the time of interaction of the external field. In the strong field limit
we can approximate the density number $n$  (\ref{electrico}) by the
expression 
\begin{equation}
n\approx \exp (-%
\frac{\pi }{2}\tanh (\frac{\pi }{4}\sqrt{8\xi -1})\frac{m^{2}}{eE}-\frac{\pi 
}{2}\sqrt{4(2\xi +k_{y}^{2}+k_{x}^{2})-1}).  \label{electrico2}
\end{equation}
Substituting (\ref{electrico2}) \ into (\ref{ene}) , we obtain that the
total number $N$ of particles per unit volume takes the form 
\begin{equation}
N\approx \frac{eE}{8\pi ^{4}T}(b+2)\exp (-b /2)\exp (-\frac{m^{2}\pi }{2eE%
}\tanh (\frac{b}{4})),  \label{N}
\end{equation}
where $b=\pi \sqrt{8\xi -1.}$ Result (\ref{N}) resembles the number of
particles created by a constant electric field in a Minkowski space\cite
{Grib,Nikishov}. It is worth mentioning that the number $N$ of particles per
unit volume is inversely proportional to $T^{-1}$ and vanishes as $%
T\rightarrow \infty $. The \ volume expansion of the anisotropic universe (%
\ref{a1}) is  faster than the particle creation process,  therefore $N$
becomes negligible  for large values of $T.$

The results (\ref{beta2}) , (\ref{electrico}), and (\ref{N}) show that the
anisotropic cosmological background (\ref{a1}), as well the constant
electric field, contribute to the creation of scalar particles. The
quasiclassical method gives a recipe for obtaining the positive and negative
frequency modes even when spacetime is not static and an external source is
present. The presence of the anisotropy with a constant electric field gives
place to a particle distribution that is thermal only in the asymptotic
field regime. The method and results presented in this paper could be of
help to discuss quantum effects in more realistic anisotropic cosmological
scenarios.

\acknowledgments We thank Dr. Juan Rivero for helpful discussions. This work
was supported by CONICIT under project 96000061.

\end{document}